\documentclass[a4paper,amsmath,amssymb,nofootinbib]{revtex4}
\makeatletter

\def \be {\begin{equation}}
\def \ee {\end{equation}}
\def \ba {\begin{array}}
\def \ea {\end{array}}
\def \bea{\begin{eqnarray}}
\def \eea{\end{eqnarray}}

\def \g {\gamma}
\def \G {\Gamma}
\def \d {\delta}

\def \m {\mu}
\def \n {\nu}

\def \l {\lambda}

\def \s {\sigma}

\def \r {\rho}
\def \o {\omega}

\def \t {\tau}

\def \p {\partial}
\def \f {\frac}
\def \na {\nabla}

\def \nn {\nonumber}
\def \ol {\overline}

\def \wh {\widehat}

\def \ra {\rightarrow}
\def \lra {\leftrightarrow}

\def \wt {\widetilde}

\begin{document}
\title{A modified variational principle for gravity in modified Weyl geometry}

\author{Fang-Fang Yuan}
\email{ffyuan@emails.bjut.edu.cn}
\affiliation{Institute of Theoretical Physics, Beijing University of Technology,
Beijing 100124, China}

\author{Yong-Chang Huang}
\email{ychuang@bjut.edu.cn}
\affiliation{Institute of Theoretical Physics, Beijing University of Technology,
Beijing 100124, China}


\begin{abstract}
The usual interpretation of Weyl geometry is modified in two senses. First, both the additive Weyl connection and its variation are treated as (1, 2) tensors under the action of Weyl covariant derivative.
Second, a modified covariant derivative operator is introduced which still preserves the tensor structure of the theory. With its help, the Riemann tensor in Weyl geometry can be written in a more compact form. We justify this modification in detail from several aspects and obtain some insights along the way.
By introducing some new transformation rules for the variation of tensors under the action of Weyl covariant derivative, we find a Weyl version of Palatini identity for Riemann tensor.
To derive the energy-momentum tensor and equations of motion for gravity in Weyl geometry,
one naturally applies this identity at first, and then converts the variation of additive Weyl connection to those of metric tensor and Weyl gauge field.
We also discuss possible connections to the current literature on Weyl-invariant extension of massive gravity and the variational principles in f(R) gravity.
\end{abstract}
\maketitle
\section{Introduction}	

In search of "a true infinitesimal geometry" which can only compare lengths at one infinitesimally close point, H. Weyl proposed that a spacetime manifold should only be equipped with a class of conformally equivalent metrics rather than a definite metric as in general relativity \cite{{Weyl1918}}. Once the metric is given a scale or "gauge" freedom like this, Riemannian geometry is generalized to the so-called Weyl geometry.

In one conformal class, a change of calibration or "gauge" for the metric is accompanied by a transformation of the corresponding differential 1-form. Interpreting the latter as electromagnetic field, and demanding the physics is "gauge" invariant, Weyl arrived at a gauge principle to unify the electromagnetism and general relativity. A decade later, Weyl updated this principle to its modern form \cite{Weyl1929}. As he conjectured, this eventually led to the establishment of gauge theory which unified "electricity and matter". (For reviews, see Refs. \cite{O'RS00} and \cite{Sc1111}.)

The relation between Weyl invariance and conformal invariance has been investigated for many years. In particular, the authors of Ref. \cite{IO'RSW96} suggested a correspondence between Weyl invariance in curved space and conformal invariance in flat space. The Brans-Dicke theory of gravity \cite{{BD1961},{D1961}} was also found to be relevant in this context. For recent discussions of this topic using the language of Weyl geometry, see Refs. \cite{{Romero2011},{RFP1106},{QGAM1108},{Fa1212}}. Other related works on Weyl invariance are Refs. \cite{{Sc1102},{FaFr1106},{Pe1110},{CL1205},{Fa1206},{CD'OPP1210}}. For some recent research on Weyl geometry from the mathematical point of view, see e.g. Refs. \cite{{HoNa2011},{Kala2011},{BGGV1211}}.


It has been a popular idea to give mass to the graviton. These massive gravities are still under intense investigation in recent years. The topologically massive gravity \cite{DJT1982} was connected to conformal field theory through the holographic principle \cite{{W2007},{LSS2008}}. Some new theories were proposed including New Massive Gravity \cite{BHT2009} and nonlinear massive gravity \cite{dRGT2010}. Weyl gravity has also attracted a lot of attention since the works in Refs. \cite{LP1101} and \cite{Ma1105}. Inspired by all these developments, some authors have studied the Weyl-invariant extension of New Massive Gravity \cite{DT1104}, Dirac-Born-Infeld type gravity \cite{MKS1106}, general higher curvature gravity theories \cite{TDT1201}, and topologically massive gravity \cite{DKT1209}.

In the Palatini approach to f(R) gravity (see e.g. Ref. \cite{So2009}), the field equation of the connection acquires a form which most resembles the Weyl connection in Weyl integrable geometry (with the gauge field $W_\m = \p_\m \phi, \ \phi = \f{1}{2} \ln \p_R f(R)$). Both metric and Palatini f(R) gravities have been related to the Brans-Dicke theory via conformal transformations \cite{So2008}. These facts may be connected with the works on Weyl geometry mentioned above \cite{{RFP1106},{QGAM1108}}. Recently, based on Refs. \cite{{1003.5532},{1010.4776},{1103.2743},{1201.4018}}, a biconnection variational principle has been proposed in Ref. \cite{Ta1205} and applied to f(R) gravity. If we set the second connection there to be the additive part of Weyl connection and introduce a scalar field, the corresponding generalized action would have some resemblance with the Weyl-invariant extension of Einstein-Hilbert action. These issues may deserve further research.

Compared to the case in Riemannian geometry, the Riemann tensor in Weyl geometry acquires an additive part which has a complicated dependence on the Weyl gauge field. When the action of a Weyl-invariant massive gravity or modified gravity is given, one typically needs to use the variational principle to find the energy-momentum tensor and the equation of motion for the Weyl gauge field. To do this, the Riemann tensor (in Weyl geometry) has to be inserted first. The subsequent calculation of variation usually becomes formidable, and one may have to resort to software. It is desirable to have a method to track the procedure. Especially, a Weyl version of Palatini identity would be much helpful.

We find that this object could be partially accomplished if one treats the additive Weyl connection and its variation as regular tensors, and introduces sensible transformation rules for them under the action of Weyl covariant derivative.
Interestingly, a modified derivative operator is found to be relevant. In contrast with Riemannian geometry, such a modification is not only possible but also preserves the Leibniz's rule (product rule) even when the metric tensor is involved. With its help, the Riemann tensor can be derived in an alternative way. On the other hand, to find a Weyl version of Palatini identity, we introduce some new transformation rules for the Weyl covariant derivative. Although they still have no analogues in Riemannian geometry as one expected, the expressions are quite natural and similar to those in general relativity. This makes them very easy to understand. Along with the exploration, we also reinterpret some peculiarities of Weyl geometry through the comparison with Riemannian geometry.

The organization of this paper is as follows. In the next section, we review the relevant basics of Weyl geometry and explain our idea in some detail. In Sec. \ref{pb wcd}, we take a close look at the peculiar behavior of Weyl covariant derivative, and find that it still allows a modification which is forbidden in the Riemannian case. In Sec. \ref{a tensor}, using the transformation rule of tensors under the action of the modified Weyl covariant derivative, we obtain some formulas for the additive Weyl connection. A new expression for the Riemann tensor is given in Sec. \ref{curv}, and with its help we find new ways to derive the curvature tensors. Sec. \ref{mvp} deals with the variational problem for gravity in Weyl geometry. After clarifying some technical issues, we finally arrive at a palatini-like identity for the Riemann tensor. This may provide a useful technique to find the energy-momentum tensor and the equations of motion. The conclusion and discussion can be found in Sec. \ref{last}.

Our conventions are mostly those of Ref. \cite{Carroll} adapted to Weyl geometry.

\section{Motivation}

Although the concept of Weyl geometry is based on Weyl transformations, it suffices for our purposes to regard it as a generalization of Riemannian geometry with a new connection. With the convention in Ref. \cite{O'RS00} and denoting the Weyl gauge field by $W_{\mu}$, the torsion-free Weyl connection is defined as follows
\bea
\wt \G_{\m\n}^{\l} &=& \G_{\m\n}^{\l} + W_{\m\n}^{\l} , \label{weyl} \\
\G_{\m\n}^{\l} &=& \f{1}{2} g^{\l\r}(\p_\m g_{\r\n} + \p_\n g_{\r\m}  - \p_\r g_{\mu\nu} ),\label{chris}
\\
W^{\l}_{\m\n}  &=& g^{\l\r} ( g_{\r\m} W_{\n} + g_{\r\n} W_{\m} - g_{\m\n} W_{\r} )   \nn \\
&=& \d^{\l}_{\m} W_{\n} + \d^{\l}_{\n} W_{\m} - g_{\m\n}W^{\l}. \label{adcon}
\eea
Here $\G_{\m\n}^{\l}$ is the Christoffel connection (Levi-Civita connection) of Riemannian geometry. For the expediency of exposition, we will refer to $W_{\m\n}^{\l}$ defined in Eq. (\ref{adcon}) as additive Weyl connection.

One can define the corresponding Weyl covariant derivative as
\be
\wt \na_{\r} T^{a_1...}_{b_1...}
= \na_{\r}  T^{a_1...}_{b_1...}   +  W^{a_1}_{\r\l} T^{\l a_2...}_{b_1...} + ...
 - W^{\l}_{\r b_1} T^{a_1...}_{\l b_2...} - ... \ , \label{cod}
\ee
where $\na_{\r}$ is the usual (Riemannian) covariant derivative operator defined with the Christoffel connection (\ref{chris}).
(To conform with the literature, we will not introduce an abbreviation such as WCD for the Weyl covariant derivative.)
The analogue of metricity (or metric-compatibility) condition is
\be
\wt\nabla_{\r} \, g_{\m\n}  = - 2 W_{\r} \, g_{\m\n} ,
\ee
which is usually taken as the definition of Weyl geometry. Notice that some papers use different conventions from us where both the additive Weyl connection and the (non-)metricity condition may not be the same as here.

What interests us the most is the Weyl-invariant Riemann tensor
\be
\wt{R}^{\r}_{\ \s\m\n}  = \p_{\m} \wt\G^{\r}_{\n\s} -  \p_{\n} \wt\G^{\r}_{\m\s}
+ \wt \G^{\r}_{\m\l}  \, \wt\G^{\l}_{\n\s} - \wt \G^{\r}_{\n\l}  \, \wt\G^{\l}_{\m\s}.
\ee
More explicitly, we have
\bea
 \wt{R}^{\r}_{\ \s\m\n} &=&  R^{\r}_{\ \s\m\n} + \wh{R}^{\r}_{\ \s\m\n}, \\
 \wh{R}^{\r}_{\ \s\m\n} &=& \na_{\m} W^{\r}_{\n\s} -  \na_{\n} W^{\r}_{\m\s}
+ W^{\r}_{\m\l}  \, W^{\l}_{\n\s} - W^{\r}_{\n\l}  \, W^{\l}_{\m\s}. \label{arie}
\eea
We will refer to $\wh{R}^{\r}_{\ \s\m\n}$ as additive Riemann tensor. If one inserts the definition of additive Weyl connection (\ref{adcon}) in the above equation, the expression would become a little lengthy. (For the detailed formula, see Eq. (\ref{rie ex}) in Sec. \ref{curv}.)

When studying gravity theories in Weyl geometry, one usually needs to deal with the tedious calculations involving the curvature tensors. One motivation of our work is to develop a useful toolkit for this. It turns out that our modification also leads to some interesting observations about the Weyl geometry. With the comparison with Riemannian geometry in mind, all our procedures and techniques are actually quite easy to understand for general relativity practitioners.

Our first observation is that the tensor property of the additive Weyl connection (\ref{adcon}) can be promoted from Riemannian geometry to Weyl geometry. From the point of view of Riemannian geometry, it is of course a plain fact that the additive Weyl connection is a (1, 2) tensor, and obeys the basic transformation rule under the action of Riemannian covariant derivative. One may wonder what happens if it is treated as a (1, 2) tensor in Weyl geometry. The behavior of the additive Weyl connection under general coordinate transformations does not concern us, instead we will concentrate on the action of Weyl covariant derivative (\ref{cod}) on it.

Following this suggestion, we have
\be
\wt \na_{\m} W^{\r}_{\n\s}
= \na_{\m}  W^{\r}_{\n\s}   +  W^{\r}_{\m\l} W^{\l}_{\n\s} - W^{\l}_{\m\n} W^{\r}_{\l\s} - W^{\l}_{\m\s} W^{\r}_{\n\l}.
\ee
Then the additive Riemann tensor (\ref{arie}) becomes
\be
\wh{R}^{\r}_{\ \s\m\n} = \wt\na_{\m} W^{\r}_{\n\s} -  \wt\na_{\n} W^{\r}_{\m\s}
- W^{\r}_{\m\l}  \, W^{\l}_{\n\s} + W^{\r}_{\n\l}  \, W^{\l}_{\m\s},
\ee
which is even harder to handle. However, if we introduce a modified Weyl covariant derivative as follows
\be
\wh \na_{\r} T^{a_1...}_{b_1...}
= \na_{\r}  T^{a_1...}_{b_1...}   +  \f{1}{2} W^{a_1}_{\r\l} T^{\l a_2...}_{b_1...} + ...
 - \f{1}{2} W^{\l}_{\r b_1} T^{a_1...}_{\l b_2...} - ... \ ,  \label{mcod}
\ee
the expression 
turns out to be more compact
\be
\wh{R}^{\r}_{\ \s\m\n} = \wh\na_{\m} W^{\r}_{\n\s} -  \wh\na_{\n} W^{\r}_{\m\s}.  \label{ma rie}
\ee
In this case, we have a modified (non-)metricity condition
\be
\wh \na_{\r} \, g_{\m\n}  = - W_{\r} \, g_{\m\n} .  \label{m met}
\ee
This will not be an issue if one notices that just like Riemannian geometry, the Weyl geometry is also completely determined by the corresponding Riemann tensor.

Let us recall the Palatini identity in Riemannian geometry (or more properly, in general relativity):
\be
\d R^{\r}_{\ \s\m\n} = \na_{\m} \d \G^{\r}_{\n\s} -  \na_{\n} \d  \G^{\r}_{\m\s}. \label{rpa}
\ee
One may be tempted to interpret the additive Weyl connection as a special variation of Christoffel connection, i.e.
$\d \G^{\r}_{\n\s} \ra W^{\r}_{\n\s}, \  \d R^{\r}_{\ \s\m\n} \ra \wh R^{\r}_{\ \s\m\n}.$ However, to pass from Eq.(\ref{rpa}) to Eq.(\ref{ma rie}), one still needs to introduce a new Weyl covariant derivative defined in Eq. (\ref{mcod}). This interesting fact is also a sign that our choice may be the unique and nontrivial one.

Although this is a pleasant result, for the modified Weyl covariant derivative to be sensible, one needs to demonstrate that under its action the tensor structure is still preserved. Since the metric-compatibility condition fails in Weyl geometry, one must make sure that the new definition is consistent with the Leibniz's rule (product rule) especially when the metric tensor is involved.
This also guarantees that the two facets of additive Weyl connection will not lead to any contradiction: besides the new character as a (1, 2) tensor (in Weyl geometry), it is still a combination of Weyl gauge fields defined in Eq. (\ref{adcon}). The detailed discussion can be found in the next two sections. We will also show that this modification has no analogue in Riemannian geometry and it is not equivalent to the case where one rescales the definition of additive Weyl connection.




The second observation is related to the variational problem for gravity in Weyl geometry. Inspired by the above idea, we find that if the variation of additive Weyl connection 
is also treated as a (1, 2) tensor, one would have a Weyl version of Palatini identity. The explicit formula is
\bea
\d\wh{R}^{\r}_{\ \s\m\n} &=& (\, \wt\na_{\m} \d + \d_{\na_{\m}} \,) W^{\r}_{\n\s} - (\, \wt \na_{\n} \d + \d_{\na_{\n}} \,) W^{\r}_{\m\s}.
\eea
This modified variational approach turns out to be quite transparent. 
However, there are some important issues we would like to elaborate on in Sec. \ref{mvp}. The explanation of the transformation rule and the delta-like operator $\d_{\na_{\m}}$ can also be found there.
While the usual approach deals with the explicit expression of Riemann tensor directly, with
the above two observations combined together we arrive at a systematic and equivalent method for the variational
problem.

At the end of this section, we would like to collect some formulas about the additive Weyl connection:
\bea
W^{\l}_{\m\l} &=& n W_{\m}, \label{for 1}  \\
g^{\m\n} W^{\l}_{\m\n} &=& -(n-2) W^{\l}, \label{for 2}  \\
W^{\n}_{\m\l} W^{\l} &=& \d^{\n}_{\m} W^2, \label{for 3}  \\
W^{\l}_{\m\n} W_{\l} &=& 2 W_{\m} W_{\n} - g_{\m\n} W^2, \label{for 4}  \\
\na_{\l} W^{\l}_{\m\n} &=& \na_{\m} W_{\n} + \na_{\n} W_{\m} - g_{\m\n} \na \cdot W. \label{for 5}
\eea
Here $n$ denotes the dimension of the differentiable manifold, i.e. the spacetime dimension, and $\na\cdot W\equiv \na_\mu W^\mu$. The derivations of these formulas are elementary. Although they are usually not explicitly spelled out in the literature, we will find them very useful for our exposition.

\section{The peculiar behavior of Weyl covariant derivatives} \label{pb wcd}

In contrast with Riemann-Cartan geometry which has non-metricity tensor $Q_{\r\m\n} \equiv  - \na_{\r} g_{\m\n} = 0$, this metric-compatibility condition fails in Weyl geometry. So the operation of raising and lowering of indices no longer commutes with the covariant derivative.

Because of this, the modified Weyl covariant derivative (\ref{mcod}) shares a peculiarity with the usual one (\ref{cod}):
 \be  \wh\na_{\m} V^{\m}  \neq  \wh\na^{\m} V_{\m},    \label{n cdot}    \ee
where $V^\m$ or $V_\m$ is a general vector. This fact can be shown as follows. For the LHS, we have
\bea
\wh\na_{\m} V^{\m} &=& \na \cdot V + \f{1}{2} W^{\m}_{\m\l} V^{\l}   \nn  \\
          &=&  \na \cdot V + \f{1}{2} n W \cdot V .
\eea
Here Eq. (\ref{for 1}) has been used, and
$W \cdot V \equiv W_{\l} V^{\l} = W^{\l} V_{\l}$ .

For the RHS, we have
\bea
\wh\na^{\m} V_{\m} &\equiv&  g^{\m\n} \wh\na_{\n} V_{\m} \nn \\
       &=&   g^{\m\n} ( \na_{\n} V_{\m} - \f{1}{2} W^{\l}_{\n\m} V_{\l})  \nn \\
       &=&   \na \cdot V + \f{1}{2} (n - 2) W \cdot V,
\eea
where Eq. (\ref{for 2}) has been used.
Another way to calculate 
it is
\bea
\wh\na^{\m} V_{\m} &=& \na \cdot V  - \f{1}{2} W^{\l\m}_{\ \ \ \m} V_{\l} \nn \\
&=& \na \cdot V  - \f{1}{2} ( g^{\l\m} W_{\m} + \d^{\l}_{\m} W^{\m} - \d^{\m}_{\m} W^{\l} ) V_{\l} \nn \\
                   &=&  \na \cdot V + \f{1}{2} (n - 2) W \cdot V,
\eea
which involves an unfamiliar additive Weyl connection (see Eq. (\ref{nadcon 2}) below). Now one naturally obtains the statement (\ref{n cdot}). This also implies that we cannot define a symbol like $\wh\na \cdot V$. Although the derivations here may seem simple, we give the details just to show how our new Weyl covariant derivative works.

Notice that the nonequality (\ref{n cdot}) is actually a special case of the following general fact
\be  \wh\na^{\m} V_{\r} \equiv  g^{\m\n} \wh\na_{\n} V_{\r} \neq \wh\na_{\n} (g^{\m\n}V_{\r}).
\ee
In spite of the above peculiarity, if we treat the metric as a usual tensor and stick to the basic transformation rule, all the calculations would still give sensible results. This is because the Weyl covariant derivatives preserve the Leibniz's rule even when the metric tensor is involved. When taking the action of them on the multiplication of any two (or more than two) tensors, one could use the Leibniz's rule first to expand it and then apply the transformation rule corresponding to the derivative operators, or reverse the order. These two ways of expansions should be consistent with each other.

To see this point more clearly, let us introduce a generalized Weyl covariant derivative as follows
\be
\ol \na_{\r} T^{a_1...}_{b_1...}
= \na_{\r}  T^{a_1...}_{b_1...}   +  s \, W^{a_1}_{\r\l} T^{\l a_2...}_{b_1...} + ...
 - t \, W^{\l}_{\r b_1} T^{a_1...}_{\l b_2...} - ... \, . \label{g cod}
\ee
Here the upper and lower "covariant weights" $s$ and $t$ are just some numbers. Notice that
\be \wt \na_{\r} \equiv \ol \na_{\r} (s=t=1), \quad \wh \na_{\r} \equiv \ol \na_{\r} (s=t=\f{1}{2}) . \label{g cod f} \ee

For ordinary tensors, the transformation rule (\ref{g cod}) always preserves the Leibniz's rule. This can be seen by expanding e.g. $\ol \na_{\r} (\, V^\m S_{\n\t} \,)$ in two ways. More explicitly, we have
\bea
\ol \na_{\r} (\, V^\m S_{\n\t} \,) &=& \ol \na_{\r} V^\m S_{\n\t}  + V^\m \ol \na_{\r} S_{\n\t}  \nn  \\
  &=& (\, \na_{\r} V^\m + s \, W^{\m}_{\r\l} V^\l \,) S_{\n\t}  \nn  \\
  &&+ V^\m (\, \na_{\r} S_{\n\t} - t \, W^{\l}_{\r\n} S_{\l\t} - t \, W^{\l}_{\r\t} S_{\n\l}  \,).
\eea
On the other hand,
\bea
\ol \na_{\r} (\, V^\m S_{\n\t} \,) &=& \na_{\r} (\, V^\m S_{\n\t} \,) + s \, W^{\m}_{\r\l} (\, V^\l  S_{\n\t} \,) \nn \\
 &&- t \, W^{\l}_{\r\n} (\,V^\m S_{\l\t} \,) - t \, W^{\l}_{\r\t} (\,V^\m S_{\n\l} \,) .
\eea
These two equations are obviously equal to each other.

When the metric tensor is involved, one should be more careful. This is essentially because it is a special tensor we use to raise or lower tensor indices.
It suffices to look at a typical example such as
\bea
{\ol\na}_\r ( g_{\m\n} V^{\n} ) &=& {\ol\na}_\r \, g_{\m\n} V^{\n} +  g_{\m\n} \ol\na_{\r} V^{\n} \nn  \\
                              &=& - 2 t \, W_\r \, g_{\m\n} V^{\n}
                              + g_{\m\n} (\, \na_{\r} V^{\n} + s \, W^{\n}_{\r\l} V^{\l} \,)    \nn  \\
                              &=& - 2 t \, W_\r V_{\m} + \na_{\r} V_{\m} - V^\n \na_\r g_{\m\n}
                                + s \, (\, g_{\r\m} W \cdot V + W_\r V_\m - W_\m V_\r   \,) \nn \\
 &=&  \na_\r V_{\m} - V^\n \na_\r g_{\m\n} + (s - 2 t) \, W_\r V_\m - s \, W_\m V_\r + s \, g_{\r\m} W \cdot V . \label{st 1}
\eea
On the other hand, we have
\bea
\ol\na_\r V_{\m} &=& \na_{\r} V_{\m} - t \, W^{\l}_{\r\m} V_{\l}  \nn  \\
             &=& \na_{\r} V_{\m} - t \, (\,  W_\r V_\m + W_\m V_\r - g_{\r\m} W \cdot V \,) . \label{st 2}
\eea

Since here $\na_\r g_{\m\n} = 0$, for these two equations to be equal to each other, a necessary condition is $s = t$. Both the usual and modified Weyl covariant derivatives satisfy this condition.
If we further require the additive Riemann tensor to take a compact form like Eq. (\ref{ma rie}), one is naturally led back to our modified derivative operator: $\wh \na_{\r} \equiv \ol \na_{\r} (s=t=\f{1}{2})$.

Although the Weyl connection itself is not metric-compatible, one can show that any connection which is not metric-compatible cannot be promoted to Weyl geometry. Accordingly there will be no sensible definition of Weyl covariant derivative to give the non-metricity condition which is the basic definition of Weyl geometry. In other words, a geometry (space) which is not Riemann-Cartan cannot be promoted to Weyl geometry.
This can be easily seen by looking at Eqs. (\ref{st 1}) and (\ref{st 2}), with $ \na_\r g_{\m\n} =0 \ra \na_\r g_{\m\n} \neq 0 $. It is interesting that one can get this expected result through such an elementary way.

Notice that in the definition of the modified Weyl covariant derivative (\ref{mcod}), we do not rescale the operator $\na_\r$ simultaneously. This is related to the following fact: In Riemannian geometry, there is no sensible definition of a generalized covariant derivative.
Suppose one is given such a definition as
\bea
\boldsymbol{\na}_{\r} T^{a_1...}_{b_1...}
= \p_{\r}  T^{a_1...}_{b_1...}   +  \boldsymbol{s} \, \G^{a_1}_{\r\l} T^{\l a_2...}_{b_1...} + ...
 - \boldsymbol t \, \G^{\l}_{\r b_1} T^{a_1...}_{\l b_2...} - ... \, .
\eea
Then we can take its action on the metric tensor and use the expression of Christoffel connection in Eq. (\ref{chris}) to arrive at the following expressions
\bea
\boldsymbol{\na}_{\r} \,  g_{\m\n} &=& \p_{\r} g_{\m\n} - \boldsymbol t \, (\, \G^{\l}_{\r \m} g_{\l\n}
                                 +  \, \G^{\l}_{\r \n} g_{\m\l} \,) \nn \\
                                   &=& (1 - \boldsymbol t) \ \p_{\r} g_{\m\n} ,   \\
\boldsymbol{\na}_{\r} \, g^{\m\n} &=& \p_{\r} g^{\m\n} + \boldsymbol s \, (\, \G^{\m}_{\r \l} g^{\l\n}
                                 +  \, \G^{\n}_{\r \l} g^{\m\l} \,)  \nn  \\
  &=&  \p_{\r} g^{\m\n} + \f{1}{2}  \boldsymbol s \, (\, \p_\r g_{\s\l} + \p_\l g_{\s\r} - \p_\s g_{\r\l} \,)
  (\, g^{\m\s} g^{\l\n} + g^{\n\s} g^{\l\m}    \,)  \nn   \\
  &\equiv&   \p_{\r} g^{\m\n} + \f{1}{2}  \boldsymbol s \,  \p_\r g_{\s\l}   (\, g^{\m\s} g^{\l\n} + g^{\n\s} g^{\l\m}  \,)  \nn  \\
  &=& \p_{\r} g^{\m\n} + \f{1}{2}  \boldsymbol s \, [\, 2 \p_{\r} g^{\m\n} - g_{\s\l} \p_ \r (\, g^{\m\s} g^{\l\n} + g^{\n\s} g^{\l\m}  \,) \,] \nn  \\
  &=& (1 - \boldsymbol s) \ \p_{\r} g^{\m\n} .
\eea
For the metricity condition $\boldsymbol{\na}_{\r} \,  g_{\m\n} = \boldsymbol{\na}_{\r} \, g^{\m\n} = 0$ to be preserved, the only choice is obviously $\boldsymbol s = \boldsymbol t = 1$.



Due to the symmetric property of the additive Weyl connection, the operators $\wh \na_{\r}$ and $\wt \na_{\r}$
actually have other good behaviors. For example, one finds that
\be
\wh\na_{\m} V^{2m}  =  \wt\na_{\m} V^{2m} =  \na_{\m} V^{2m} .
\ee
If we define the Weyl gauge field strengths as usual
\bea
F_{\m\n} &\equiv& \na_{\m} W_{\n} - \na_{\n} W_{\m } , \label{wfs 1} \\
\wt F_{\m\n} &\equiv& \wt\na_{\m} W_{\n} - \wt\na_{\n} W_{\m } ,  \label{wfs 2} \\
\wh F_{\m\n} &\equiv& \wh\na_{\m} W_{\n} - \wh\na_{\n} W_{\m } ,  \label{wfs 3}
\eea
one can also check the following fact
\be
\wh F_{\m\n} = \wt F_{\m\n} = F_{\m\n}.  \label{wfs}
\ee

\section{Additive Weyl connection as a tensor in Weyl geometry} \label{a tensor}

The central task of this section is to check the action of modified Weyl covariant derivative (\ref{mcod}) on the additive Weyl connection (\ref{adcon}), and show that it does not spoil the tensor property of the latter. The procedure and formulas here are the basis of the next section where we will focus on the curvature tensors in (modified) Weyl geometry.


\subsection{A preliminary discussion} \label{pre}

It is a trivial fact that the additive Weyl connection is a (1, 2) tensor in Riemannian geometry. What we would like to find is its behavior under the action of the usual or modified Weyl covariant derivative. Before doing this, we need to clarify two special issues.

Firstly, since we do not introduce a Weyl-rescaled metric tensor like $g'_{\m\n} = e^{2\o} g_{\m\n}$ here, one can raise or lower the indices of the additive Weyl connection with the same metric tensor as in Riemannian geometry. Specifically, we can define
\bea
W_{\l,\m\n}  &=&  g_{\l\m} W_{\n} + g_{\l\n} W_{\m} - g_{\m\n} W_{\l}, \label{nadcon 1} \\
W^{\l\m}_{\ \ \ \n}  &=&  g^{\l\m} W_{\n} + \d^{\l}_{\n} W^{\m} - \d^{\m}_{\n} W^{\l}, \label{nadcon 2} \\
{W_{\l}}^{\m}_{\ \n}  &=&  \d^{\m}_{\l} W_{\n} + g_{\l\n} W^{\m} -\d^{\m}_{\n} W_{\l},
\eea
and others. (We have added a comma in Eq. (\ref{nadcon 1}) to stress its symmetric property.) They all obey the regular transformation rule under the action of Riemannian covariant derivative. However, these tensor properties of $W^{\l}_{\m\n}$ have nothing to do with the Weyl geometry per se.

Secondly, our modification by introducing a new Weyl covariant derivative (\ref{mcod}) is not equivalent to the case where one rescales the definition of the additive Weyl connection (\ref{adcon}) while retaining the usual Weyl covariant derivative (\ref{cod}). Suppose we have a new additive Weyl connection as follows
\bea
\ol{W} &=& W,  \\
\ol{W}^{\l}_{\m\n} &=& \f{1}{2} ( \d^{\l}_{\m} \ol{W}_{\n} + \d^{\l}_{\n} \ol{W}_{\m} - g_{\m\n} \ol{W}^{\l} ) \nn \\
 &=& \f{1}{2} {W}^{\l}_{\m\n}.
\eea
Demanding the corresponding Weyl covariant derivative to behave as the usual one in (\ref{cod}), one has
\bea
\ol\na_{\r} V^{\m} &=& \na_{\r} V^{\m} + \ol{W}^{\m}_{\r\l} V^{\l} \nn \\
&=& \na_{\r} V^{\m} + \f{1}{2}{W}^{\m}_{\r\l} V^{\l}, \\
\ol\na_{\r} g_{\m\n} &=&  \na_{\r} g_{\m\n} - \ol{W}^{\l}_{\r\m} g_{\l\n} - \ol{W}^{\l}_{\r\n} g_{\m\l}  \nn  \\
&=&  - W_{\r} g_{\m\n},
\eea
which are the same as our modified case. Nevertheless, for the additive Riemann tensor (\ref{arie}), one would have
\bea
\wh{\ol{R}}^{\ \r}_{\ \s\m\n} &=& \na_{\m} \ol{W}^{\r}_{\n\s} -  \na_{\n} \ol{W}^{\r}_{\m\s}
+ \ol{W}^{\r}_{\m\l}  \, \ol{W}^{\l}_{\n\s} - \ol{W}^{\r}_{\n\l}  \, \ol{W}^{\l}_{\m\s}  \nn \\
&=&  \f{1}{2} ( \na_{\m} W^{\r}_{\n\s} -  \na_{\n} W^{\r}_{\m\s} )
+ \f{1}{4} ( W^{\r}_{\m\l}  \, W^{\l}_{\n\s} - W^{\r}_{\n\l}  \, W^{\l}_{\m\s} ).
\eea
This is clearly different from our case. One could also try to rescale the Weyl gauge field while retaining the definition of additive Weyl connection. This would still change the value of the Riemann tensor.

\subsection{Some formulas involving the additive Weyl connection}

Before proceeding further, let us encapsulate our modification of the Weyl geometry as follows. The Weyl connection is kept unchanged as in Eqs. (\ref{weyl}) - (\ref{adcon})
while we introduce a modified Weyl covariant derivative in Eq. (\ref{mcod}). Accordingly the non-metricity condition
is modified to Eq. (\ref{m met}).


Using the definition in Eq. (\ref{mcod}), we can easily find the following basic formulas for the Weyl gauge field
\bea
\wh \na_{\r} W^{\m} &=& \na_{\r} W^{\m} + \f{1}{2} W^{\m}_{\r\l} W^{\l}  \nn \\
&=&  \na_{\r} W^{\m} + \f{1}{2} \d^{\m}_{\r} W^2  , \label{wg 1} \\
\wh \na_{\r} W_{\m} &=& \na_{\r} W_{\m} - \f{1}{2} W^{\l}_{\r\m} W_{\l}  \nn \\
&=& (\na_{\r} - W_{\r}) W_{\m} + \f{1}{2} g_{\r\m} W^2 , \label{wg 2}
\eea
where Eqs. (\ref{for 3}) and (\ref{for 4}) have been used.
Another two useful equations are
\bea
\wh\na_{\r} W^{\r} &=&  \na \cdot W + \f{1}{2} n W^2, \label{wg 3}  \\
\wh\na^{\r} W_{\r} &=&   \na \cdot W + \f{1}{2} (n - 2) W^2.  \label{wg 4}
\eea
Notice that similar equations also appeared in Sec. \ref{pb wcd}.

Now we can study the behavior of additive Weyl connection $W^{\l}_{\m\n}$ under the action of our modified Weyl covariant derivative. Inserting the definition of $W^{\l}_{\m\n}$, we have
\bea
\wh\na_{\r} W^{\l}_{\m\n}
&\equiv& \wh \na_{\r} ( \d^{\l}_{\m} W_{\n} + \d^{\l}_{\n} W_{\m} - g_{\m\n}W^{\l} ) \nn \\
 &=& \d^{\l}_{\m} \wh\na_{\r} W_{\n} + \d^{\l}_{\n} \wh\na_{\r} W_{\m} - g_{\m\n} \wh\na_{\r} W^{\l}
  + g_{\m\n} W_{\r} W^{\l}, \label {na w 1}
\eea
where the modified non-metricity condition (\ref{m met}) has been used.

With the basic formulas (\ref{wg 1}) and (\ref{wg 2}) at hand, one can expand the above equation as
\bea
 \wh\na_{\r} W^{\l}_{\m\n}
 &=&   \d^{\l}_{\m} [ \, (\na_{\r} - W_{\r}) W_{\n} + \f{1}{2} g_{\r\n} W^2 \, ]
    +  \d^{\l}_{\n} [ \, (\na_{\r} - W_{\r}) W_{\m} + \f{1}{2} g_{\r\m} W^2 \, ]  \nn  \\
    &&-  g_{\m\n} (\na_{\r} W^{\l} + \f{1}{2} \d^{\l}_{\r} W^2 ) + g_{\m\n} W_{\r} W^{\l}  \nn \\
    &=& ( \na_{\r} - W_{\r}) (\d^{\l}_{\m} W_{\n} + \d^{\l}_{\n} W_{\m} - g_{\m\n}W^{\l})  \nn  \\
     &&+ \f{1}{2} ( \d^{\l}_{\m} g_{\r\n} + \d^{\l}_{\n} g_{\r\m} - \d^{\l}_{\r} g_{\m\n} ) W^2 \nn  \\
    &=& ( \na_{\r} - W_{\r}) W^{\l}_{\m\n} + \f{1}{2} ( \d^{\l}_{\m} g_{\r\n} + \d^{\l}_{\n} g_{\r\m} - \d^{\l}_{\r} g_{\m\n} ) W^2 .
\eea
This is an important equation which will be used to obtain the explicit expressions for curvature tensors in Sec. \ref{curv}.

Since we have asserted that $W^{\l}_{\m\n}$ can be treated as a (1, 2) tensor in Weyl geometry, one should check this statement by doing the calculation in another way. Using the transformation rule in Eq. (\ref{mcod}), we have
\be
\wh\na_{\r} W^{\l}_{\m\n} = \na_{\r} W^{\l}_{\m\n} + \f{1}{2} W^{\l}_{\r \s} W^{\s}_{\m \n}
                            - \f{1}{2} W^{\s}_{\r \m} W^{\l}_{\s \n} - \f{1}{2} W^{\s}_{\r \n} W^{\l}_{\m \s} .  \label{ba W}
\ee
Inserting the following equation
\be
W^{\l}_{\r \s} W^{\s}_{\m \n}
= \d^{\l}_{\r} W_{\s} W^{\s}_{\m \n} +  W_{\r} W^{\l}_{\m \n} - g_{\r\s}W^{\l} W^{\s}_{\m \n},
\ee
one is led to a quite lengthy calculation
\bea
\wh\na_{\r} W^{\l}_{\m\n} &= \na_{\r} W^{\l}_{\m\n} &+ \f{1}{2} [ \, W^{\l}_{\r \s} W^{\s}_{\m \n}
           -  (\r \lra \n)  -  (\r \lra \m) \, ]   \nn  \\
           &=   \na_{\r} W^{\l}_{\m\n} &+ \f{1}{2} [ \, \d^{\l}_{\r} W_{\s} W^{\s}_{\m \n} -  (\r \lra \n)  -  (\r \lra \m) \, ]  \nn  \\
                   &&+ \f{1}{2} [ \,  W_{\r} W^{\l}_{\m \n}  -  (\r \lra \n)  -  (\r \lra \m)  \, ]   \nn  \\
                   &&- \f{1}{2} [ \,  g_{\r\s}W^{\l} W^{\s}_{\m \n}  -  (\r \lra \n)  -  (\r \lra \m)  \, ]  \nn  \\
 &=  \na_{\r} W^{\l}_{\m\n} &+ \f{1}{2} [ \, 2 \d^{\l}_{\r} W_{\m} W_{\n}
     - 2 W_{\r} ( \d^{\l}_{\m} W_{\n}+\d^{\l}_{\n} W_{\m})
          +  (\d^{\l}_{\m} g_{\r\n} + \d^{\l}_{\n} g_{\r\m} - \d^{\l}_{\r} g_{\m\n}) W^2 \, ] \nn  \\
          &&+ \f{1}{2} [\,  -2 \d^{\l}_{\r} W_{\m} W_{\n} + W^{\l} (g_{\r\m}W_{\n} + g_{\r\n}W_{\m} - g_{\m\n}W_{\r}) \, ] \nn \\
         && - \f{1}{2} W^{\l} [\, g_{\r\m} W_{\n}  + g_{\r\n} W_{\m} - 3 g_{\m\n} W_{\r}   \,]  \nn  \\
 &=  ( \na_{\r} - W_{\r}&) W^{\l}_{\m\n} + \f{1}{2} ( \d^{\l}_{\m} g_{\r\n} + \d^{\l}_{\n} g_{\r\m} - \d^{\l}_{\r} g_{\m\n} ) W^2 .
\eea
Thus these two procedures are consistent with each other.

Let us rewrite this important formula below
\be
\wh\na_{\r} W^{\l}_{\m\n} = ( \na_{\r} - W_{\r}) W^{\l}_{\m\n} + \f{1}{2} ( \d^{\l}_{\m} g_{\r\n} + \d^{\l}_{\n} g_{\r\m} - \d^{\l}_{\r} g_{\m\n} ) W^2 .  \label{na w}
\ee
With its help, we will derive the expressions of curvature tensors in the next section.

\section{Curvature tensors in modified Weyl geometry} \label{curv}

When a modified Weyl covariant derivative is introduced as in Eq. (\ref{mcod}), we can rephrase the additive Riemann tensor (\ref{arie}) in a more compact form
\be
\wh{R}^{\r}_{\ \s\m\n} = \wh\na_{\m} W^{\r}_{\n\s} -  \wh\na_{\n} W^{\r}_{\m\s}.  \label{ma rie 2}
\ee
This can be seen by recalling the basic equation in (\ref{ba W}).
In what follows, we will use this new formula to obtain the explicit expressions for Riemann tensor, Ricci tensor, and Ricci scalar in (modified) Weyl geometry successively. The results can also be checked to agree with the usual approach.

(i) Riemann tensor     \\
Inserting Eq. (\ref{na w 1}) in the new formula (\ref{ma rie 2}), we immediately arrive at
\bea
\wh{R}^{\r}_{\ \s\m\n} = \d^{\r}_{\s} \wh F_{\m\n} - 2 \d^{\r}_{[\m} \wh \na_{\n]} W_{\s}
                       + 2 ( g_{\s[\m} \wh \na_{\n]} -  g_{\s[\m} W_{\n]} )  W^{\r} .   \label{m rie}
\eea
Our convention here is $A_{[\m\n]} \equiv \f{1}{2} ( A_{\m\n} - A_{\n\m} ) $. The field strength $\wh F_{\m\n}$ is defined as 
in Eq. (\ref{wfs 3}).

Using the basic formulas Eqs. (\ref{wg 1}) and (\ref{wg 2}), and noticing the fact $\wh F_{\m\n} = F_{\m\n}$ (see Eq. (\ref{wfs})), we reproduce the well-known Weyl-invariant Riemann tensor
\bea
\wt{R}^{\r}_{\ \s\m\n} &=&  R^{\r}_{\ \s\m\n} + \wh{R}^{\r}_{\ \s\m\n},  \nn  \\
\wh{R}^{\r}_{\ \s\m\n} &=&  \d^{\r}_{\s} F_{\m\n} - 2 \d^{\r}_{[\m} \na_{\n]} W_{\s}
                                   + 2 g_{\s[\m} \na_{\n]} W^{\r}   \nn   \\
                                   &&- 2 W_{[\m} \d^{\r}_{\n]} W_{\s} - 2 g_{\s[\m} W_{\n]} W^{\r}
                                   + 2 g_{\s[\m} \d^{\r}_{\n]} W^2  .  \label{rie ex}
\eea

We can also use Eq. (\ref{na w}) instead and insert it in Eq. (\ref{ma rie 2}) to arrive at
\be
\wh{R}^{\r}_{\ \s\m\n} = \na_{\m} W^{\r}_{\n\s} -  \na_{\n} W^{\r}_{\m\s}
- W_{\m} W^{\r}_{\n\s} + W_{\n} W^{\r}_{\m\s}  +  ( \d^{\r}_{\n} g_{\m\s} - \d^{\r}_{\m} g_{\n\s} ) W^2 .
\ee
With the definition of $W^{\l}_{\m\n}$, one expands the first four terms as
\bea
\na_{\m} W^{\r}_{\n\s} -  \na_{\n} W^{\r}_{\m\s} &=& \d^{\r}_{\s} F_{\m\n} - 2 \d^{\r}_{[\m} \na_{\n]} W_{\s}
                                   + 2 g_{\s[\m} \na_{\n]} W^{\r},  \\
W_{\m} W^{\r}_{\n\s} - W_{\n} W^{\r}_{\m\s}  &=& 2 W_{[\m} \d^{\r}_{\n]} W_{\s} + 2 g_{\s[\m} W_{\n]} W^{\s}.
\eea
Finally, we still get the result in Eq. (\ref{rie ex}).

(ii) Ricci tensor   \\
Instead of contracting the indices in the explicit formula for Riemann tensor, let us use the key formula (\ref{ma rie 2}) directly to find the Weyl-invariant Ricci tensor. Now we have
\bea
\wh R_{\s\n} &=& \wh\na_{\r} W^{\r}_{\n\s} - \wh\na_{\n} W^{\r}_{\r\s}   \nn  \\
             &=& \wh F_{\s\n} - (n-2) \wh\na_{\n} W_{\s} - g_{\s\n} ( \wh\na_{\r} W^{\r} - W^2 ). \label{m ric}
\eea
Using the basic formulas Eqs. (\ref{wg 1}) and (\ref{wg 2}) again, we get the final expression
\bea
\wt R_{\s\n} &=& R_{\s\n} +  \wh R_{\s\n}, \nn  \\
\wh R_{\s\n} &=& F_{\s\n} - (n-2) [\, (\na_{\n} - W_{\n}) W_{\s} + g_{\s\n} W^2 \,] - g_{\s\n} \na \cdot W . \label{ric}
\eea

There is still another way to do the calculation, i.e. using the formula (\ref{na w}). Contracting the indices there, we have
\bea
\wh\na_{\r} W^{\r}_{\n\s} &=& \na_{\r} W^{\r}_{\n\s} - W_{\r} W^{\r}_{\n\s} - \f{1}{2} (n-2) g_{\n\s} W^2 \nn  \\
                          &=& (\na_{\s} W_{\n} + \na_{\n} W_{\s} - g_{\s\n} \na \cdot W)
                             - (2 W_{\s} W_{\n} - g_{\s\n} W^2) \nn  \\
                             && - \f{1}{2} (n-2)  g_{\s\n}  W^2 ,
\eea
where Eqs. (\ref{for 4}) and (\ref{for 5}) have been used.
Subtracting it by another term
\be
\wh\na_{\n} W^{\r}_{\r\s} = n (\na_{\n} - W_{\n}) W_{\s} + \f{1}{2} n\, g_{\s\n} W^2 ,
\ee
one arrives at the result in Eq. (\ref{ric}) again.

Notice that the Ricci tensor in Eq. (\ref{ric}) is not symmetric. It is customary to split it into two parts as
\bea
\wh R_{\s\n} &=& \wh R_{[\s\n]} +  \wh R_{<\s\n>},   \\
\wh R_{[\s\n]} &=& F_{\s\n} ,  \\
\wh R_{<\s\n>} &=& - (n-2) [\, (\na_{\n} - W_{\n}) W_{\s} + g_{\s\n} W^2 \,] - g_{\s\n} \na \cdot W .
\eea

(iii) Ricci scalar    \\
Using the definition of Ricci scalar in Weyl geometry and Eq. (\ref{m ric}), we have
\bea
\wh R &\equiv& g^{\s\n} \wh R_{\s\n}      \nn  \\
      &=& - (n-2) \wh\na^{\r} W_{\r} - n ( \wh\na_{\r} W^{\r} - W^2 ).   \label{m rics}
\eea
Inserting Eqs. (\ref{wg 3}) and (\ref{wg 4}) in the above equation, we arrive at
\bea
\wt R &=& R  + \wh R , \nn  \\
\wh R &=& - 2 (n-1) \na \cdot W - (n-1)(n-2) W^2 .
\eea

Notice that the Ricci scalar here is not Weyl-invariant: under the transformation $g'_{\m\n} = e^{2\o} g_{\m\n}$, it behaves like $\wt R' = e^{- 2\o} \wt R$. We need to introduce a compensating scalar field to construct a Weyl-invariant extension of Einstein-Hilbert action
\be
S = \int d^n x \sqrt{-g} ( \Phi^2 \wt R + ...  ) \ .
\ee

The Einstein tensor has the following form
\bea
\wt G_{\m\n}  &=&  G_{\m\n} + \wh G_{\m\n}  , \nn \\
\wh G_{\m\n}  &=&  \wh R_{<\m\n>} - \f{1}{2} g_{\m\n} \wh R  \nn \\
              &=&  (n-2) \{\ - (\na_{\n} - W_\n) W_\m + g_{\m\n} [ \na \cdot W + \f{1}{2} (n-3) W^2 ] \ \} .
\eea

All the above curvature tensors we have obtained in modified Weyl geometry (Eqs. (\ref{m rie}), (\ref{m ric}) and (\ref{m rics})) are consistent with the usual approach. (One may compare our results with e.g. Ref. \cite{DT1104}.) This means that we are not introducing a new geometry here, but only modify the Weyl geometry in some sense.

This can also be seen from the following fact: The Ricci and Bianchi identities are still the same as in usual Weyl geometry. More explicitly, one has
\bea
[ \wt\na_\m , \wt\na_\n ] V^\r &=& \wt{R}^{\r}_{\ \s\m\n} V^\s ,   \\
\wt{R}^{\r}_{\ [\s\m\n]} &=& 0 , \\
\wt\na_{[\l} \wt{R}^{\ \ \ \ \r}_{\n\m]\s} &=& 0 .
\eea
One should not be tempted to write $ [ \wh\na_\m , \wh\na_\n ] V^\r \bumpeq \wh{R}^{\r}_{\ \s\m\n} V^\s $, $ [ \wh\na_\m , \wh\na_\n ] V^\r \bumpeq \wt{R}^{\r}_{\ \s\m\n} V^\s $, etc. Here we use the symbol "$\bumpeq$" to stress that these are questionable formulas.
All in all, we actually keep the mathematical structure of Weyl geometry intact.

\section{A modified variational principle}  \label{mvp}

\subsection{Main obstacles}

Given a gravity action, one usually needs to vary it with respect to the metric to get the energy-momentum tensor. As for its Weyl-invariant extension, one also needs to find the equation of motion for the Weyl gauge field through variation. A key formula for this variational principle is the Palatini identity in Riemannian geometry (or more properly, in general relativity):
\be
\d R^{\r}_{\ \s\m\n} = \na_{\m} \d \G^{\r}_{\n\s} -  \na_{\n} \d  \G^{\r}_{\m\s}. \label{rpa 2}
\ee

It is desirable to have a Weyl version of this identity for the additive Riemann tensor. Although we have already obtained a compact form for it in Eq. (\ref{ma rie 2}), to find an analogous formula for its variation turns out to a quite nontrivial job. This is because of the following problem:
\bea
\d \wh{R}^{\r}_{\ \s\m\n} \equiv \d \wh\na_{\m} W^{\r}_{\n\s} -  \d \wh\na_{\n} W^{\r}_{\m\s}
                          &\neq& \wh\na_{\m} \d W^{\r}_{\n\s} -  \wh\na_{\n} \d W^{\r}_{\m\s}  \nn  \\
                          &=&  \wh\na_{\m} (\, \d^{\r}_{\n} \, \d W_\s + ...   \,) - ...    \nn  \\
                          &\circeq&  \d^{\r}_{\n} (\, \na_{\m} \d W_\s - \f{1}{2} W^{\l}_{\m\s} \d W_\l \,) + ... \, .
\eea
Here we use the symbol "$\circeq$" to stress that an unverified transformation rule has been applied.
The problem remains even we choose another procedure as
\bea
&&\wh\na_{\m} \d W^{\r}_{\n\s} -  \wh\na_{\n} \d W^{\r}_{\m\s}  \nn  \\
&\circeq& \na_{\m} \d W^{\r}_{\n\s} + \f{1}{2} W^{\r}_{\m\l} \d W^{\l}_{\n\s} + ... \, .
\eea
In the above calculations we vary $W^{\r}_{\n\s}$ with respect to $g_{\g\d}$ and $W_\t$ at the same time. The situation is unchanged when one does the variations individually. Also in the practical calculation, one usually uses the integration by parts to write $\Phi^2 \na_\r \d g_{\m\n}$ as $- \na_\r \Phi^2 \d g_{\m\n}$. These technical issues do not concern us here.

We conclude that the symbol of variation does not commute with the Weyl covariant derivative operator(s). The above results can be checked by hand. Since the calculations are very lengthy and non-illuminating, we will not report them here.

Things become more interesting when one finds that even the Riemannian covariant derivative does not commute with the symbol of variation when acting on the additive Weyl connection:
\be
\d \na_{\m} W^{\r}_{\n\s} \neq \na_{\m} \d W^{\r}_{\n\s}. \label{neq w}
\ee

As we will show, all these problems can be solved if we notice a peculiarity inherited from Riemannian geometry, and define a sensible transformation rule for the variation of general tensors in Weyl geometry. Since the usual Weyl covariant derivative becomes more relevant here rather than our modified one, the following mnemonic is useful: in the language of Eqs. (\ref{g cod}) and (\ref{g cod f}), their "covariant weights" are $1$ and $\f{1}{2}$, respectively.

\subsection{Basic definitions}

The phenomenon in (\ref{neq w}) can be easily explained if one notices the following proposition: In Riemann-Cartan geometry, when the involved composite tensor has an explicit dependence on the metric, the symbol of variation does not commute with the Riemannian covariant derivative. This is because of a very simple fact:
\be
[\,\d \, , \na_{\r}\,] \, g_{\m\n} = - \na_{\r} \, \d  g_{\m\n},
\ee
where the metricity condition $\na_{\r} g_{\m\n} = 0$ has been used. In other words, we have
\be
[\,\d \, , \na_{\r}\,] \ (... + g_{\m\n} S^{a_{1}...}_{b_{1}...} +...) \neq 0.  \label{neq r}
\ee
Notice that the tensor $S^{a_{1}...}_{b_{1}...}$ here should have no common indices with $g_{\m\n}$. Otherwise one would have a rather unpleasant result: $[\,\d \, , \na_{\r}\,] \ V_\m \equiv 0$ while $[\,\d \, , \na_{\r}\,] \ ( g_{\m\n}V^\n ) \neq 0$.

We then assume that any tensor and its variation obey exactly the same transformation rule under the action of the usual Weyl covariant derivative as follows
\be
\wt\na_{\r} \, \d  T^{a_{1}...}_{b_{1}...} = \na_{\r} \, \d T^{a_{1}...}_{b_{1}...} +  W^{a_1}_{\r\l} \d T^{\l a_2...}_{b_1...} + ...
 - W^{\l}_{\r b_1} \d T^{a_1...}_{\l b_2...} - ... \ . \label{dcod}
\ee
It should be noted that this definition has no analogue in Riemannian geometry (or more properly, in general relativity). This means that one cannot define a similar transformation rule such as
\be
\na_{\r} \, \d  T^{a_{1}...}_{b_{1}...} \circeq \, \p_{\r} \d T^{a_{1}...}_{b_{1}...} +  \G^{a_1}_{\r\l} \d T^{\l a_2...}_{b_1...} + ...
 - \G^{\l}_{\r b_1} \d T^{a_1...}_{\l b_2...} - ... \ . \label{n rcd}
\ee
It would be a disaster for general relativists since we always write $\d \na_\r T^{a_{1}...}_{b_{1}...} = \na_\r \d T^{a_{1}...}_{b_{1}...}$ for a single tensor. For the LHS, we have
\be
\d \na_\r  T^{a_{1}...}_{b_{1}...} =\p_{\r} \d T^{a_{1}...}_{b_{1}...} + \d (\G^{a_1}_{\r\l}  T^{\l a_2...}_{b_1...}) + ...
 - \d (\G^{\l}_{\r b_1} T^{a_1...}_{\l b_2...}) - ... \ . \label{n rcd l}
\ee
These two equations are clearly not equal to each other.

However, when it comes to the additive Weyl connection,
our definition indeed has strong resemblance with the transformation rule for the variation of Christoffel connection
\bea
\wt\na_{\r} \, \d W^{\l}_{\m\n} &=& \na_{\r} \, \d W^{\l}_{\m\n} +  W^{\l}_{\r\s} \d W^{\s}_{\m\n}
 - W^{\s}_{\r \m} \d W^{\l}_{\s\n} - W^{\s}_{\r \n} \d W^{\l}_{\m\s}\   , \\
\na_{\r} \, \d \G^{\l}_{\m\n} &=& \p_{\r} \, \d \G^{\l}_{\m\n} +  \G^{\l}_{\r\s} \d\G^{\s}_{\m\n}
 - \G^{\s}_{\r \m} \d\G^{\l}_{\s\n} - \G^{\s}_{\r \n} \d\G^{\l}_{\m\s}\ . \label{na chr}
\eea
While the Christoffel connection $\G^{\l}_{\m\n}$ itself is not a tensor and cannot be given a sensible transformation rule under the action of covariant derivative, its variation $C^{\l}_{\m\n} \equiv \d \G^{\l}_{\m\n}$ is a genuine tensor. So Eq. (\ref{na chr}) is not a manifestation of the questionable rule in Eq. (\ref{n rcd}). In contrast, we treat both the additive Weyl connection $W^{\l}_{\m\n}$ and its variation $\d W^{\l}_{\m\n}$ as (1, 2) tensors in Weyl geometry, and demand them to obey exactly the same transformation rule (see Eqs. (\ref{cod}) and (\ref{dcod})).

In view of the fact that the symbol of variation does not commute with the Weyl covariant derivative, we introduce another definition
\be
[\,\d \, , \wt\na_{\r}\,] T^{a_1...}_{b_1...} = \d W^{a_1}_{\r\l} T^{\l a_2...}_{b_1...} + ...
 - \d W^{\l}_{\r b_1} T^{a_1...}_{\l b_2...} - ... + [\,\d \, , \na_{\r}\,] T^{a_1...}_{b_1...}. \label{com}
\ee
It also has no analogue in Riemannian geometry. By inspection of Eqs. (\ref{n rcd}) and (\ref{n rcd l}), one may be tempted to introduce a similar definition
\be
[\,\d \, , \na_{\r}\,] T^{a_1...}_{b_1...} \circeq \d \G^{a_1}_{\r\l} T^{\l a_2...}_{b_1...} + ...
 - \d \G^{\l}_{\r b_1} T^{a_1...}_{\l b_2...} - ... \ . \label{n rcd 2}
\ee
However, with Eqs. (\ref{n rcd}) and (\ref{n rcd 2}) combined together, all the calculations involving variation in general relativity would still be substantially changed. The mathematical structure there may be spoiled because of this. One can also check that our definitions have no conflict with Riemannian geometry by simply setting $W^{\l}_{\m\n} \ra 0$ in them. In one word, we do not change anything about the Riemannian covariant derivative in this work.

With the above definitions (\ref{dcod}) and (\ref{com}) at hand, we finally arrive at the following identity for the variation of Riemann tensor in Weyl geometry
\bea
\d \wt R^{\r}_{\ \s\m\n} &=& \d R^{\r}_{\ \s\m\n} + \d \wh R^{\r}_{\ \s\m\n}, \nn  \\
\d \wh R^{\r}_{\ \s\m\n} &=& (\, \wt\na_\m \d + [\d , \na_\m] \,) W^{\r}_{\n\s}
                         - (\, \wt\na_\n \d + [\d , \na_\n] \,) W^{\r}_{\m\s} .  \label{main}
\eea
This identity is not complicated as it appears, since $\d_{\na_\m} \equiv [\d , \na_\m]$ is just a delta-like operator, and is always trivial unless acting on the metric tensor. In other words, all we need is
\be
[\d , \na_\m] \ W^{\r}_{\n\s} = \na_{\m} \, \d  g_{\n\s} \, W^\r.  \label{c awc}
\ee

One may try to modify the definition (\ref{dcod}) to the following unfamiliar form
\be
\wt{\ol\na}_{\r} \, \d  T^{a_{1}...}_{b_{1}...} = \d \, \na_{\r} T^{a_{1}...}_{b_{1}...} +  W^{a_1}_{\r\l} \d T^{\l a_2...}_{b_1...} + ...
 - W^{\l}_{\r b_1} \d T^{a_1...}_{\l b_2...} - ... \ . \label{n dcod}
\ee
Then we will have a more pleasant result
\bea
\d \wh R^{\r}_{\ \s\m\n} &=& \, \wt{\ol\na}_\m \d  W^{\r}_{\n\s}
                         - \, \wt{\ol\na}_\n \d  W^{\r}_{\m\s} .  \label{main 2}
\eea
Also we can totally forget about the peculiarity concerning the operator $\d_{\na_\m} \equiv [\d , \na_\m]$, which is just an unfortunate heritage from Riemannian geometry.

\subsection{The variation of Ricci tensor}

In the following, we will use the Weyl version of Palatini identity (\ref{main}) to find the variation of the additive Ricci tensor. Contracting the indices there and using Eq. (\ref{c awc}), we have
\bea
\d \wh R_{\s\n} &=&  (\, \wt\na_\r \d + [\d , \na_\r] \,) W^{\r}_{\n\s}
                         - (\, \wt\na_\n \d + [\d , \na_\n] \,) W^{\r}_{\r\s}   \nn  \\
&=& \wt\na_\r \d W^{\r}_{\n\s} + W \cdot \na \d g_{\n\s} - n \wt\na_\n \d W_\s .    \label{v ric 1}
\eea
Notice that due to the remark under Eq. (\ref{neq r}), $0 = [\d , \na_\n] W^{\r}_{\r\s} \neq \na_{\n} \, \d  g_{\r\s} \, W^\r$. If one uses Eq. (\ref{main 2}) instead (as we recommend), then this peculiarity can be totally dismissed.

The first term in the above equation can be expanded as
\bea
\wt\na_\r \d W^{\r}_{\n\s} &=& \wt\na_\r (\,  \d^{\r}_{\n} \d W_{\s} + \d^{\r}_{\s} \d W_{\n} - g_{\n\s} \d W^{\r} - W^\r \d g_{\n\s} \,)  \nn  \\
&=& \wt \na_\n \d W_{\s} + \wt \na_\s \d W_{\n} - \wt \na_\r g_{\n\s} \d W^{\r} - g_{\n\s} \wt \na_\r \d W^{\r} \nn \\
&&-\wt \na_\r  W^\r \d g_{\n\s} - W^\r \wt \na_\r \d g_{\n\s}  \nn  \\
&=& \na_\s \d W_{\n} + \na_\n \d W_{\s} - 2 W^{\r}_{\s\n} \d W_\r - g_{\s\n} [\, \na_\r + (n-2) W_\r \,] \d W^\r  \nn \\
&&- [\, \na \cdot W +  W \cdot \na + (n-2) W^2 \,] \d g_{\s\n} . \label{term 1}
\eea
Here we have used the following basic formulas
\bea
\wt\na_\s \d W_\n &=& \na_\s \d W_\n - W^{\r}_{\s\n} \d W_\r , \\
\wt\na_\r \d W^\r &=& (\, \na_\r + n W_\r \,) \d W^\r  ,  \\
W^\r \wt\na_\r \d g_{\n\s} &=& (\,W \cdot \na  - 2 W^2 \,) \d g_{\n\s} .
\eea
There is of course another way to do the calculation in Eq. (\ref{term 1}) which is a little lengthy
\bea
\wt\na_\r \d W^{\r}_{\n\s} &=& \na_\r \d W^{\r}_{\n\s} + W^{\r}_{\r\l} \d W^{\l}_{\n\s} - W^{\l}_{\r\n} \d W^{\r}_{\l\s} - W^{\l}_{\r\s} \d W^{\r}_{\n\l}   \nn   \\
&=& \na_\n \d W_\s + \na_\s \d W_\n - g_{\n\s} \na_\r \d W^\r
                       - (\, \na \cdot W + W \cdot \na \,) \d g_{\n\s} \nn  \\
    && + n \, (\, W_\s \d W_\n + W_\n \d W_\s  - g_{\n\s} W_\l \d W^\l  -  W^2 \d g_{\n\s}    \,)  \nn  \\
    && - (\, n \, W_\n \d W_\s + W^{\l}_{\s\n} \d W_\l - g_{\l\s} W^{\l}_{\r\n} \d W^\r -  W^2 \d g_{\n\s}       \,)  \nn  \\
    && - (\, W^{\l}_{\n\s} \d W_\l + n \,  W_\s \d W_\n - g_{\n\l} W^{\l}_{\r\s} \d W^\r -  W^2 \d g_{\n\s}    \,)     \nn  \\
 &=&    \na_\s \d W_{\n} + \na_\n \d W_{\s} - 2 (\, W_\s \d W_\n + W_\n \d W_\s  \,) \nn \\
 && - g_{\s\n} [\, \na_\r + (n-4) W_\r \,] \d W^\r \nn  \\
 &&- [\, \na \cdot W +  W \cdot \na + (n-2) W^2 \,] \d g_{\s\n} .  \label{term 1 a}
\eea
The result is the same as expected.

Inserting Eq. ({\ref{term 1}}) or Eq. ({\ref{term 1 a}}) in Eq. (\ref{v ric 1}), the variation of the additive Ricci tensor is obtained as follows
\bea
\d \wh R_{\s\n}
          &=& \na_\s \d W_\n - (n-1) \na_\n \d W_\s + (n-2) (\,  W_\s \d W_\n + W_\n \d W_\s \,) \nn  \\
          &&- g_{\s\n} [\, \na_\r + 2(n-2) W_\r  \,]  \d W^\r - [\, \na \cdot W + (n-2) W^2 \,] \d g_{\s\n} . \label{ric va}
\eea
From this, one can get another useful equation
\bea
g^{\s\n} \d \wh R_{\s\n} &\equiv& g^{\s\n} \d \wh R_{<\s\n>}   \nn  \\
               &=& - 2(n-1) [\, \na_\r  + (n-2) W_\r \,] \d W^\r   \nn  \\
                          && - g^{\s\n} [\, \na \cdot W + (n-2) W^2 \,] \d g_{\s\n} .
\eea
Notice that the above derivations require no knowledge of the explicit form of Weyl-invariant Ricci tensor as in Eq. (\ref{ric}) or Eq. (\ref{m ric}).

Analogous definitions as in Eqs. (\ref{dcod}) and (\ref{com}) can also be given for the modified Weyl covariant derivative. To check this, let us start from the additive Ricci tensor in Eq. (\ref{m ric}) and take its variation. Then we will have
\bea
\d \wh R_{\s\n} &=& \d \wh F_{\s\n} - (n-2) \d \wh \na_\n W_\s - (\wh \na_\r W^\r - W^2) \d g_{\s\n} - g_{\s\n} \d \wh \na_\r W^\r + 2 g_{\s\n} W_\r \d W^\r     \nn  \\
&=& \wh \na_\s \d W_\n - \wh \na_\n \d W_\s - (n-2) (\wh \na_\n \d W_\s - \f{1}{2} \d W^{\r}_{\n\s} W_\r)
- (\na \cdot W + \f{1}{2} n W^2 \nn  \\
&& - W^2) \d g_{\s\n} - g_{\s\n} (\wh \na_\r \d W^\r + \f{1}{2} \d W^{\r}_{\r\l} W^\l) + 2 g_{\s\n} W_\r \d W^\r \nn  \\
&=& \na_\s \d W_\n - \na_\n \d W_\s - (n-2) (\na_\n \d W_\s - \f{1}{2} W^{\r}_{\n\s} \d W_\r - \f{1}{2} \d W^{\r}_{\n\s} W_\r)   \nn  \\
&&- (\na \cdot W + \f{1}{2} n W^2 - W^2) \d g_{\s\n} - g_{\s\n} (\na_\r \d W^\r + \f{1}{2} W^{\r}_{\r\l} \d W^\l + \f{1}{2} \d W^{\r}_{\r\l} W^\l)  \nn  \\
&&+ 2 g_{\s\n} W_\r \d W^\r .
\eea
Using the equation (see Eq. (\ref{for 4}))
\be
\d (W^{\r}_{\n\s} W_\r) 
= 2 (W_\n \d W_\s + W_\s \d W_\n - g_{\n\s} W_\r \d W^\r - \f{1}{2} W^2 \d g_{\n\s}),
\ee
and collecting all the terms, one still arrives at the result in Eq. (\ref{ric va}).

Although all these procedures give the consistent results, in practical calculation we choose the following way: use the Weyl version of Palatini identity in Eq. (\ref{main}), and insert the definition of additive Weyl connection, then apply the transformation rule in Eq. (\ref{dcod}) to expand the expression as in Eq. (\ref{term 1}). The final result is the same as the usual approach which takes the variation of the explicit expressions of curvature tensors directly.

To apply our method to the Weyl-invariant extension of higher curvature gravity theories or other complicated situations (see Refs. \cite{{DT1104},{MKS1106},{TDT1201},{DKT1209}}), one still needs more work. Other issues may also become relevant. We find it more suitable to return to this whole topic in another occasion.

\section{Conclusion}    \label{last}

Our main results can be summarized as follows. If we treat both the additive Weyl connection and its variation as ordinary (1, 2) tensors, and demand them to obey similar transformation rules under the action of Weyl covariant derivatives, then the Riemann tensor and its variation could be written in compact forms.

In addition to the usual approach which deals with the calculations involving curvature tensors by using the explicit expressions of the latter directly, now we have two new ways to obtain exactly the same results. One is free to choose between the following alternative procedures: (i) use the (covariant) transformation rule of additive Weyl connection as a (1, 2) tensor, and then insert the definition of the latter; (ii) insert its definition at first, and then use the transformation rule of Weyl gauge field as a vector.

For the convenience of the reader, we list the basic formulas below.

(i) A generalized Weyl covariant derivative is defined as
\be
\ol \na_{\r} T^{a_1...}_{b_1...}
= \na_{\r}  T^{a_1...}_{b_1...}   +  s \ W^{a_1}_{\r\l} T^{\l a_2...}_{b_1...} + ...
 - t \ W^{\l}_{\r b_1} T^{a_1...}_{\l b_2...} - ... \, . \label{g cod 2}
\ee
Two necessary conditions for it to sensible are $\na_{\r} g_{\m\n} = 0$, and $s=t$.
The usual Weyl covariant derivative and the modified one have "covariant weights" 1 and $\f{1}{2}$, respectively:
$ \wt \na_{\r} \equiv \ol \na_{\r} (s=t=1), \quad \wh \na_{\r} \equiv \ol \na_{\r} (s=t=\f{1}{2})$.

(ii) Any tensor and its variation obey similar transformation rules under the action of the usual Weyl covariant derivative. This means that we have the following definition
\be
\wt \na_{\r} \d T^{a_1...}_{b_1...}
= \na_{\r}  \d T^{a_1...}_{b_1...}   +  W^{a_1}_{\r\l} \d T^{\l a_2...}_{b_1...} + ...
 - W^{\l}_{\r b_1} \d T^{a_1...}_{\l b_2...} - ... \ .
\ee
As remarked in the text, this definition has no analogue in Riemannian geometry.

(iii) The symbol of variation and the Weyl covariant derivative do not commute, but we can consistently define
\be
(\, [\,\d \, , \wt\na_{\r}\,] - [\,\d \, , \na_{\r}\,] \,)T^{a_1...}_{b_1...} = \d W^{a_1}_{\r\l} T^{\l a_2...}_{b_1...} + ...
 - \d W^{\l}_{\r b_1} T^{a_1...}_{\l b_2...} - ... \, .
\ee
Here $[\,\d \, , \na_{\r}\,]$ is just a delta-like operator. (One can define $\d_{\na_\r} \equiv [\,\d \, , \na_{\r}\,]$.) It is always trivial except when acting on a composite tensor which has an explicit dependence on the metric. This is because of a simple fact:
\be
[\,\d \, , \na_{\r}\,] \, g_{\m\n} = - \na_{\r} \, \d  g_{\m\n}.
\ee

(iv) The Riemann tensor and its variation in Weyl geometry can be written in the following compact forms
\bea
\wt R^{\r}_{\ \s\m\n} &=& R^{\r}_{\s\m\n} +  \wh R^{\r}_{\s\m\n} ,   \\
\wh R^{\r}_{\ \s\m\n} &=& \wh \na_\m W^{\r}_{\n\s} - \wh \na_\n W^{\r}_{\m\s}  ,  \label{a rie f}  \\
\d \wt R^{\r}_{\ \s\m\n} &=& \d R^{\r}_{\s\m\n} +  \d \wh R^{\r}_{\s\m\n} ,   \\
\d \wh R^{\r}_{\ \s\m\n} &=& (\, \wt\na_\m \d + [\d , \na_\m] \,) W^{\r}_{\n\s}
                         - (\, \wt\na_\n \d + [\d , \na_\n] \,) W^{\r}_{\m\s} . \label{a vrie f}
\eea
Here the definition of the additive Weyl connection is $W^{\l}_{\m\n}  = \d^{\l}_{\m} W_{\n} + \d^{\l}_{\n} W_{\m} - g_{\m\n}W^{\l}$. With Eqs. (\ref{a rie f}) and (\ref{a vrie f}) combined together, we arrive at a new and systematic approach to deal with all the calculations involving curvature tensors for gravity theories in Weyl geometry.

Some comments are in order here. (i) We systematically absorb the Christoffel connection into the Riemannian covariant derivative, so most of our results can be easily extended to the case with torsion, i.e. Riemann-Cartan-Weyl geometry. In that case, the Riemann tensor would have an extra term $2 \G ^{\l}_{[\m\n]} W^{\r}_{\l\s}$. (ii) In our new definitions, nothing about the Riemannian covariant derivative is changed. Even the peculiarity concerning the delta-like operator is just a plain fact although it may be overlooked by some authors. (iii) In Riemannian geometry, we could not define a generalized covariant derivative because of the metricity condition. Although in Weyl geometry this condition is relaxed, it does not mean that one can define an arbitrary covariant derivative. (iv) In the process of our exploration, we have encountered some general propositions. Although they may be commonplace to the literature, it is still interesting to interpret them in new ways.

As we have remarked in the introduction, the field equation of the connection in Palatini f(R) gravity is much like the Weyl connection in Weyl integrable geometry (see e.g. Ref. \cite{So2009}). The Palatini formalism used there also has some resemblance with the situation in Weyl geometry. So it would be interesting to find more connections between these two areas.
However, one should be aware of the differences there, e.g. the works on variational principles in f(R) gravity naturally took non-metricity tensor and torsion into consideration.
The biconnection variational principle proposed in Ref. \cite{Ta1205} is also interesting. Nevertheless, if the second connection is replaced by the additive Weyl connection, the difference tensor used there will not be directly applicable to the case of Weyl geometry. On the other hand, the corresponding action is still not in exactly the same form as the Weyl-invariant extension of Einstein-Hilbert action.
In spite of these problems, it seems that the relation between f(R) gravity and the Brans-Dicke theory could be rephrased in the language of Weyl geometry. All these issues may need more investigation to be clarified.

To apply the Weyl version of Palatini identity to explicit examples as in Refs. \cite{{DT1104},{MKS1106},{TDT1201},{DKT1209}}, one has to deal with the operation of raising and lowering of indices, the mixing of Riemann tensor in Riemannian geometry and its additive extension, and other technical issues. The nonsymmetric property of Ricci tensor in Weyl geometry may also lead to new problems. It would be nice to work the details out and see if any new technique is needed. One could also consider the Weyl-invariant extension of actions involving the covariant derivatives of curvature tensors which may have not been studied in the literature.
Finally, from the phenomenological point of view the Weyl integrable case deserves more attention.

\begin{acknowledgments}
We would like to thank Taeyoon Moon for correspondence.
We also thank an anonymous referee for an insightful suggestion regarding the viability of de Rham-Gabadadze-Tolley massive gravity.
This work was supported by National Natural Science Foundation of
China (No. 11275017 and No. 11173028).
\end{acknowledgments}

\appendix


\end{document}